\documentclass[a4paper]{article}

\usepackage{LaTeX/INTERSPEECH2022}

\usepackage[utf8]{inputenc}
\usepackage{kotex}

\usepackage[noadjust]{cite} 

\usepackage{multirow}
\usepackage{booktabs}

\usepackage{amsmath}
\usepackage{bm}
\usepackage{amsfonts} 
\usepackage{url}

\newcommand\blfootnote[1]{%
  \begingroup
  \renewcommand\thefootnote{}\footnote{#1}%
  \addtocounter{footnote}{-1}%
  \endgroup
}

\title{Hierarchical and Multi-Scale Variational Autoencoder
\\for Diverse and Natural Non-Autoregressive Text-to-Speech}
\name{Jae-Sung Bae$^{1\dagger}$, Jinhyeok Yang$^{2\dagger}$, Tae-Jun Bak$^3$, Young-Sun Joo$^3$}
\address{$^1$Samsung Research, Republic of Korea \\
$^2$Supertone, Inc., Republic of Korea \\
$^3$AI Center, NCSOFT Corp., Republic of Korea}
\email{js3.bae@samsung.com, yangyangii@supertone.ai, \{happyjun, ysjoo555\}@ncsoft.com}

\begin{document}

\maketitle
\begin{abstract}
This paper proposes a hierarchical and multi-scale variational autoencoder-based non-autoregressive text-to-speech model (HiMuV-TTS) to generate natural speech with diverse speaking styles. Recent advances in non-autoregressive TTS (NAR-TTS) models have significantly improved the inference speed and robustness of synthesized speech. However, the diversity of speaking styles and naturalness are needed to be improved. To solve this problem, we propose the HiMuV-TTS model that first determines the global-scale prosody and then determines the local-scale prosody via conditioning on the global-scale prosody and the learned text representation. In addition, we improve the quality of speech by adopting the adversarial training technique. Experimental results verify that the proposed HiMuV-TTS model can generate more diverse and natural speech as compared to TTS models with single-scale variational autoencoders, and can represent different prosody information in each scale.
\end{abstract}
\noindent\textbf{Index Terms}: non-autoregressive text-to-speech, hierarchical variational autoencoder, multi-scale prosody modeling

\section{Introduction}
\label{sec:intro}
Text-to-speech (TTS) is a typical one-to-many (O2M) mapping problem because human speaks speech in a diverse way. 
Thanks to recent advances in deep learning techniques, the neural TTS models, especially autoregressive TTS (AR-TTS) models \cite{tacotron2, deepvoice3, jsbae_speed}, address this problem via conditioning on the previously generated acoustic features and have dramatically improved the naturalness of synthesized speech. However, because of this autoregressive architecture, these models have slow inference speed and instability in the inference phase such as skipping and repeating. Recently, non-autoregressive TTS (NAR-TTS) models have significantly improved the inference speed and robustness of synthesized speech \cite{fastspeech, fastpitch, fastspeech2, glow-tts, vits, jsbae21_interspeech, parallel-tacotron, tjbak}. However, because the NAR-TTS models cannot use previously generated acoustic features as context, the O2M problem becomes severe, which results in a performance gap between the AR-TTS models, and a lack of diversity.
\blfootnote{$^\dagger$ Work performed while at NCSOFT.}

Prosody modeling is a crucial factor in alleviating the O2M problem and improving the performance of the NAR-TTS model. In \cite{fastpitch, fastspeech2}, the authors propose the NAR-TTS models that model the variance information of speech with explicit prosody features such as pitch and energy. In the inference phase, these models can generate speech with various speaking styles by changing the predicted prosody feature values. However, because they are trained in a supervised manner utilizing hand-crafted prosody features (e.g., duration, pitch, and energy) as labels, other prosody information that is intrinsically presented in speech (e.g., emotion) may not be modeled.

Other approaches to address the O2M problem of speech are by representing the variance information of speech into latent variables in an unsupervised manner \cite{gst, reference-tacotron, robust_and_fine_grained, fine_grained_robust_amazon}. These methods can generate speech with various speaking styles by modifying the latent variables. In \cite{gst} and \cite{reference-tacotron}, the authors represent prosody in a global-scale. That is, the utterance-level speaking style of a speech is modeled as a single latent vector. On the other hand, in \cite{robust_and_fine_grained} and \cite{fine_grained_robust_amazon}, the prosody is modeled in a local-scale as a sequence of latent variables. These approaches model the locally varying prosody, such as accents or prosody that differs in phoneme-level depending on the nuance, which cannot be captured in the global-scale prosody modeling approaches. However, because of the absence of global-scale representation, the local-scale prosody models cannot generate a speech with a consistent speaking style. To compensate for these drawbacks, the multi-scale prosody representation method was proposed in \cite{multi-scale_style_control}. 
This method represents the global and local scale prosody in separate latent spaces and controls the prosody in multi-scale by providing different reference speech for each scale, respectively. 
However, to generate a speech with natural prosody, a reference speech that contains a similar text content to an input text is necessary in the inference phase.

Recently, Variational autoencoders (VAEs) \cite{vae} have achieved a promising performance in the TTS domain \cite{gmvae-tacotron, parallel-tacotron, bvae-tts, vits, fully_hierarchical_and_fine_grained}. Because VAEs are a class of probabilistic generative models, TTS models with VAE-based prosody module can generate speech in multiple ways for the same text via sampling. Therefore, the VAEs are suitable to handle the O2M problem of TTS. Furthermore, they have the advantage that reference speech is not required in the inference phase. 
In \cite{parallel-tacotron}, the authors proposed the NAR-TTS model incorporated with two different kinds of VAE modules, i.e., global and local scale VAEs, respectively. However, in our preliminary experiments, the expressiveness of the single-scale VAEs was limited. TTS models incorporated with the hierarchical VAE architectures \cite{gmvae-tacotron, fully_hierarchical_and_fine_grained} have also been widely studied to improve the expressiveness of VAE-based TTS models, however, they have been studied with the AR-TTS models.

To improve the naturalness and expressiveness in NAR-TTS models, several previous approaches adopted the hierarchical structure \cite{jsbae21_interspeech, hie_2_slt, hie_3_apple}. In our previous work \cite{jsbae21_interspeech}, we showed that adopting a proper hierarchical structure is important for the NAR-TTS models. However, we have focused on improving the naturalness and intelligibility of the NAR-TTS models instead of prosody modeling. In \cite{hie_2_slt} and \cite{hie_3_apple}, the authors proposed the NAR-TTS model that adopts the explicit prosody module with a hierarchical structure. VQ-VAE \cite{vqvae} was used in \cite{hie_2_slt} to better model the prosody but they only focused on the local-scale prosody modeling. In \cite{hie_3_apple}, the authors build a hierarchical structure by using utterance-level prosody predictors in addition to phoneme-level prosody predictors. However, as in \cite{fastpitch, fastspeech2}, only the pre-defined prosody features are modeled in a supervised manner.

Taking advantage of the upper mentioned prosody modeling approaches, we propose an NAR-TTS model with hierarchical and multi-scale VAE (HiMuV-TTS) to better address the O2M problem in the NAR-TTS models. We conjecture that the prosody of speech should be modeled in multiple scales with a hierarchical structure. Global-scale prosody such as overall pitch and speaking speed should be determined first, and then local-scale prosody such as intonation within a word, duration of a phoneme, and pauses between words should be modeled depending on it. Accordingly, the HiMuV-TTS model first extracts the global-scale prosody embedding from the mel-spectrogram. Then, it is replicated and concatenated with the learned text representation and aligned with the mel-spectrogram to form local-scale prosody embeddings. Finally, the prosody embeddings of both scales are concatenated and provided as a condition to the decoder. To further improve the speech quality, we adopted the adversarial training technique in \cite{ganspeech}. In the experiments, we verified that the HiMuV-TTS model generates more diverse and natural speech compared to single-scale VAE-based TTS models. In addition, we showed that the different prosody information can be represented in each scale of the HiMuV-TTS model.

\begin{figure*}[t]
\centering
{
    \centering
    \begin{minipage}[t]{.49\linewidth}
      \centering
      \centerline{\includegraphics[width=0.9\linewidth]{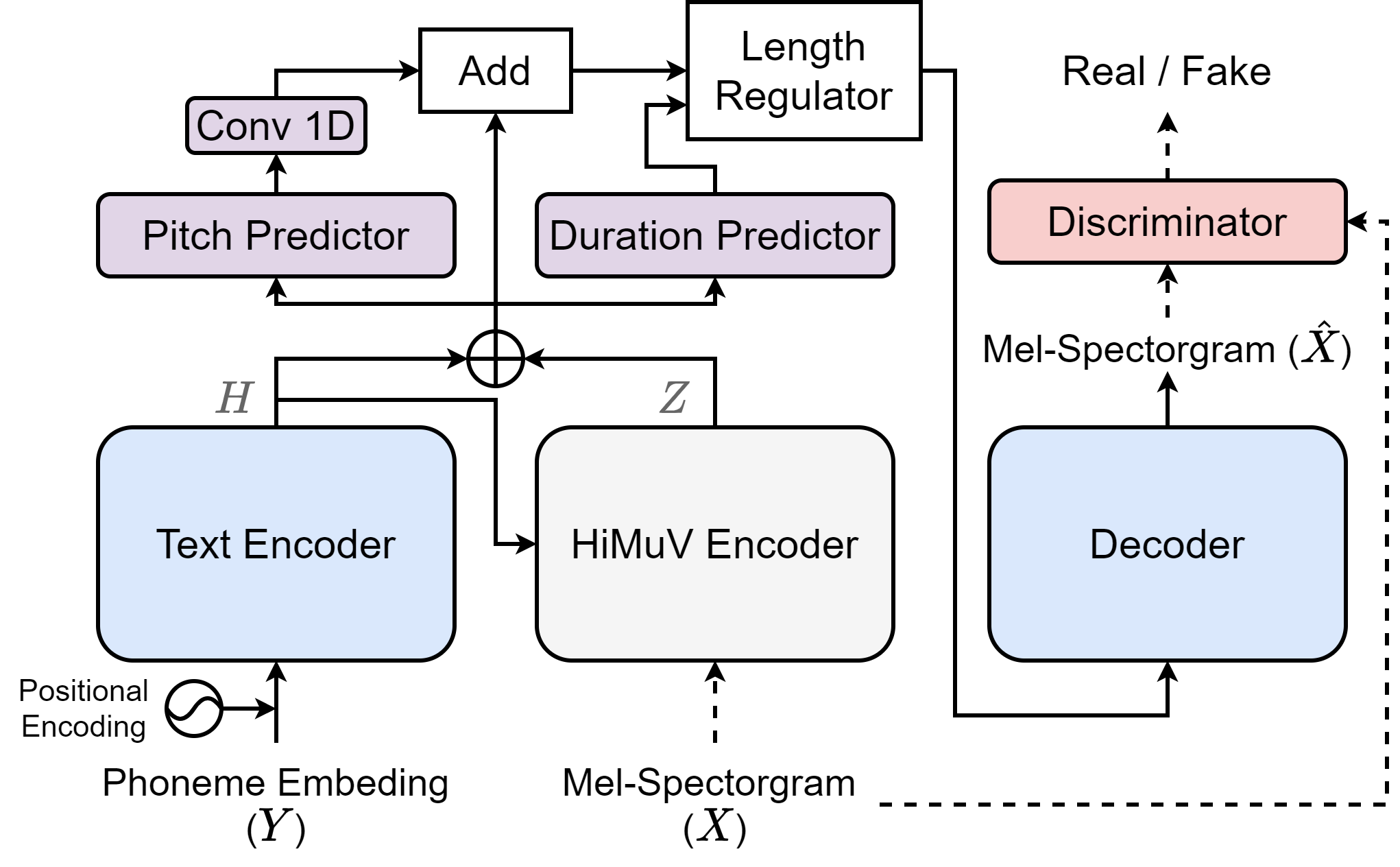}}
      \centerline{(a)} 
    \end{minipage}
    \centering
    \begin{minipage}[t]{.49\linewidth}
      \centering
      \centerline{\includegraphics[width=0.9\linewidth]{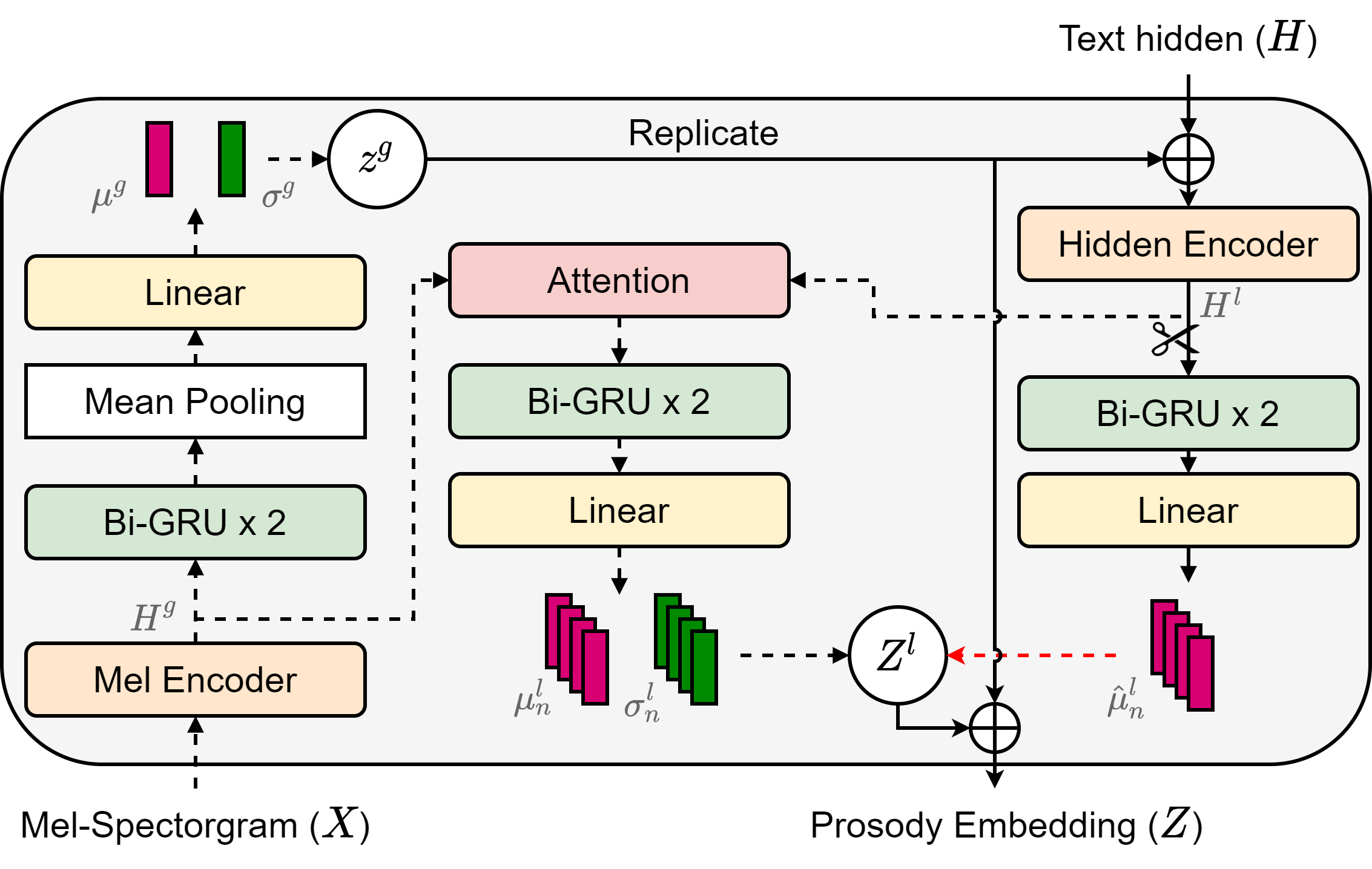}}
      \centerline{(b)}
    \end{minipage}
    \vspace{-0.2cm}
    \caption{Architecture of (a) the HiMuV-TTS model and (b) the HiMuV-encoder. The black and red dotted lines in (a) and (b) are only turned on during the training and inference phases, respectively. The black solid lines are turned on during both phases. The scissors icon in (b) indicates stop-gradient.}
    \label{fig:model}
    \vspace{-0.2cm}
}
\end{figure*}

\section{Proposed methods}
\subsection{Overall Architecture}
The overall architecture of the HiMuV-TTS model is shown in Figure \ref{fig:model}a. The input phoneme is first embedded into the phoneme embeddings $Y=\{\bm{y}_i\}_{i=1,...,N}$ and then encoded into a hidden representation $H=\{\bm{h}_i\}_{i=1,...,N}$ using a text-encoder. The proposed hierarchical and multi-scale VAE (HiMuV) encoder encodes the prosody features of target mel-spectrogram $X=\{\bm{x}_j\}_{j=1,...,M}$ into prosody embedding $Z$ conditioned on $H$. $N$ and $M$ are the number of phonemes and the number of frames of the mel-spectrogram, respectively. The duration and pitch predictors predict each feature in the phoneme-level from $H$ concatenated with $Z$. The predicted pitch values $\hat{\bm{p}}$ are embedded, added with concatenated $H$ and $Z$ and then length-regulated with the predicted duration $\hat{\bm{d}}$. Then, the decoder predicts the mel-spectrogram $\hat{X}$ from the given inputs. Finally, the discriminator distinguishes the fake/truth of $X$ and $\hat{X}$ and improves the perceptual quality of the generated speech. The reconstruction loss of the predicted mel-spectrogram, duration, and pitch values are as follows:
\begin{equation}
    L_{recon} = f(X, \hat{X}) + \alpha f(\bm{d}, \hat{\bm{d}}) + \alpha f(\bm{p}, \hat{\bm{p}}),
\end{equation}
where $\bm{d}$, $\bm{p}$, $\alpha$, and $f(\cdot)$ are the target duration, target pitch, predictor loss weight, and mean square error function, respectively.

\subsection{HiMuV Encoder}
The HiMuV-encoder first represents the global-scale prosody from the mel-spectrogram as global-scale latent variables $\bm{z}^g$ and then models the local-scale latent variables $Z^l=\{\bm{z}^l_i\}_{i=1,...,N}$ conditioned on $\bm{z}^g$ and the hidden representation $H$. 
The architecture of the HiMuV-encoder is depicted in Figure \ref{fig:model}b.
First, to model the global-scale prosody, it encodes the mel-spectrogram $X$ into the intermediate feature $H^{g}$ with the mel-encoder. After two layers of a bidirectional GRU \cite{gru} (Bi-GRU), it is averaged through the time axis, projected through a single linear layer, and become the posterior mean $\bm{\mu}^g$ and standard deviation (SD) $\bm{\sigma}^g$ of the global-scale latent variables. The global-scale latent variables $\bm{z}^g \sim \mathcal{N}(\bm{\mu}^g, ({\bm{\sigma}^g})^2)$ are then sampled from the normal distribution. $\bm{z}^g$ is then replicated and concatenated with the hidden representation $H$ and forms $H^l$ through the hidden-encoder. Next, the multi-head attention layer learns the correlation between $H^g$ and $H^l$ and outputs phoneme-level features. The two layers of Bi-GRU followed by a single linear layer predict $\bm{\mu}^l_n$ and $\bm{\sigma}^l_n$, which are the posterior mean and SD of the local-scale latent variables of the $n$-th time-step, respectively. The local-scale latent variables $\bm{z}_n^l \sim \mathcal{N}(\bm{\mu}^l_n, ({\bm{\sigma}^l_n})^2)$ is sampled from them, where $n=1,2,...,N$. Finally, the replicated $\bm{z}^g$ and $Z^l$ are concatenated and become the final prosody embedding $Z$. The prior over latent variables of both scales are assumed to follow a standard normal distribution. The model is optimized with the evidence lower bound (ELBO):
\begin{align}
    ELBO &= \mathbb{E}_{q(Z|X,Y)}\big[\log p(X|Y,Z)\big] + L_{KL} \\
    L_{KL} &= -\beta_l \sum_{n=1}^N \big[D_{KL}(q(\bm{z}^l_n|X, \bm{z}^g, H) || p(\bm{z}^l_n))\big] \nonumber \\
        &\;\;\;\; -\beta_g D_{KL}(q(\bm{z}^g|X) || p(\bm{z}^g))
\end{align}

\subsubsection{Posterior Mean Predictor}
Similar to \cite{generating_diverse_and_natural, parallel-tacotron}, we train the explicit predictor that predicts the posterior mean of local-scale latent variables to generate natural speech. The predicted values are used as the prior mean of local-scale latent variables in the inference phase. The posterior mean predictor consists of two layers of Bi-GRU followed by a single linear layer; $H^l$ is provided as an input, and it outputs the predicted posterior mean $\hat{\bm{\mu}}^l_n$ at every $n$-th time step. To prevent the posterior mean predictor from affecting other parts of the model, the stop-gradient is adopted. The posterior mean predictor is trained with the following loss function:
\begin{equation}
    L_{post} = \sum_{n=1}^N f(\bm{\mu}^l_n, \hat{\bm{\mu}}^l_n)
\end{equation}

\subsection{Discriminator and Adversarial Training}
To improve the perceptual quality of the generated speech, we adopted the adversarial training technique. The discriminator $D$ distinguishes between the predicted mel-spectrogram $\hat{X}$, which is the output of the decoder $G$, and the target mel-spectrogram $X$. The architecture of the discriminator is similar as that used in \cite{ganspeech} except that we remove the conditional part because we are building a single speaker TTS model in this work. As in \cite{ganspeech}, we used the following least-squares loss $L_{adv}$ for the adversarial training and an additional feature matching loss $L_{fm}$ for training the generator:
\begin{align}
    L_{adv}(D) =& \mathbb{E}_{(X,Y,Z)}\big[(D(X)-1)^2 + D(\hat{X})^2\big]
    \\
    L_{adv}(G) =& \mathbb{E}_{(Y,Z)}\big[(D(\hat{X})-1)^2\big]
    \\
    L_{fm}(G, D) =& \mathbb{E}_{(X,Y,Z)}\bigg[\sum_{t=1}^{T} \frac{1}{N_t}||D_t(X) - D_t(\hat{X})||_1\bigg]
\end{align}
where $T$ is the total number of layers in the discriminator and $N_t$ is the number of output features of the $t$-th layer discriminator $D_t$.

\subsection{Final Loss Function}
Equation (\ref{eq:final_loss}) is the final loss function used for training, where $\gamma$ and $\delta$ are the weights of the posterior mean prediction loss and feature matching loss, respectively.
\begin{equation}
    L_{final} = L_{recon} + L_{KL} + \gamma L_{post} + L_{adv} + \delta L_{fm}
    \label{eq:final_loss}
\end{equation}

\section{Experiments}
\subsection{Experimental Setup}
The LJ speech dataset \cite{ljspeech} was used for the experiments. It consists of 13,100 utterances from a single English female speaker with a total of approximately 24 hours. It was down-sampled to 22.05KHz. We split 131 utterances for testing and all the remaining utterances were used for training. An 80-bin mel-spectrogram with a hop size of 256 and a window size of 1024 was computed. The phoneme-level duration values were carried out using the Montreal forced alignment \cite{mfa} tool. The phoneme-level pitch values were obtained by averaging the pitch values for each phoneme. The PRAAT toolkit \cite{praat} was used to extract pitch values.

The text-encoder and decoder of the HiMuV-TTS model have the same architecture as described in \cite{fastpitch}. For the mel-encoder and the hidden-encoder of the HiMuV-encoder, the architecture used in \cite{meta-stylespeech} was adopted. The hidden dimensions of the Bi-GRU and attention layer of the HiMuV-encoder were both 128. The attention layer had two heads. The dimensions of $\bm{z}^g$ and $\bm{z}^l$ were 32 and 16, respectively. The architecture of the duration and pitch predictors consists of two layers of LSTM \cite{lstm} with 256 hidden dimensions followed by a single projection layer. The one dimensional convolutional layer with the kernel size of 3 and the stride of 1 was used for pitch embedding.

For the training, the Adam optimizer \cite{adam} was used to optimize the network with an initial learning rate of 0.002 and a batch size of 32. It was trained for 200K iterations. To avoid the collapse in VAEs, the KL-weight schedule was applied for $\beta_g$ and $\beta_l$. Each was linearly increased from 0 to $10^{-7}$ and $10^{-4}$, respectively, from 10K to 60K. For the loss weights, $\alpha$, $\gamma$, and $\delta=0.01$ were used. As a neural vocoder, VocGAN \cite{vocgan}, trained using the same speech dataset, was used.

Four models were used for the comparison; the \textbf{FastPitch} \cite{fastpitch} model, the \textbf{GANSpeech} model \cite{ganspeech}, and the global-scale VAE (\textbf{GVAE}) and the local-scale VAE (\textbf{LVAE})-based TTS models. 
The GVAE and LVAE models were chosen as single-scale VAE models. 
The VAE modules of the GVAE and LVAE models had the similar architecture as those in \cite{parallel-tacotron}, but the mel-encoder layer of the HiMuV-encoder was adopted instead of LConv layers. For the fair comparison, we changed the baseline TTS model of all comparison models to the FastPitch model. 

\begin{table}[t]
{\small
    \caption{SD values of utterance length ($\sigma_l$), average energy ($\sigma_e$), average pitch ($\sigma_p$), and pitch SD within utterance ($\sigma_{\sigma_p}$). High $\sigma_p$ value of the LVAE model results in unnatural speech (Table \ref{table:mos}).}
    \vspace{-0.6cm}
    \label{table:result1}
    \begin{center}      
    \begin{tabular}{lrrrrr}
        \toprule
        \multicolumn{1}{l}{\textbf{Model}} &
        \multicolumn{1}{c}{\textbf{$\sigma_{l}$} (s)} & 
        \multicolumn{1}{c}{\textbf{$\sigma_e$} (dB)} &
        \multicolumn{1}{c}{\textbf{$\sigma_p$} (Hz)} &
        \multicolumn{1}{c}{\textbf{$\sigma_{\sigma_p}$} (Hz)} \\
        \midrule
        GVAE & $0.31$ &  $0.75$ & $10.54$ & $8.78$ \\
        LVAE & $0.13$ &  $0.84$ & $\textbf{21.48}$ & $9.41$ \\
        HiMuV-TTS & $\textbf{0.50}$ &  $\textbf{1.02}$ & $12.01$ & $\textbf{10.44}$ \\
        \midrule
        HiMuV-TTS-G & $0.33$ & $0.86$ & $11.82$ & $10.39$ \\
        HiMuV-TTS-L & $0.36$ & $0.71$ & $2.46$ & $5.37$ \\
        \bottomrule
    \end{tabular}
    \end{center}
    \vspace{-0.5cm}
}
\end{table}

\begin{figure}[t]
    \centering
    \centerline{\includegraphics[width=0.8\linewidth]{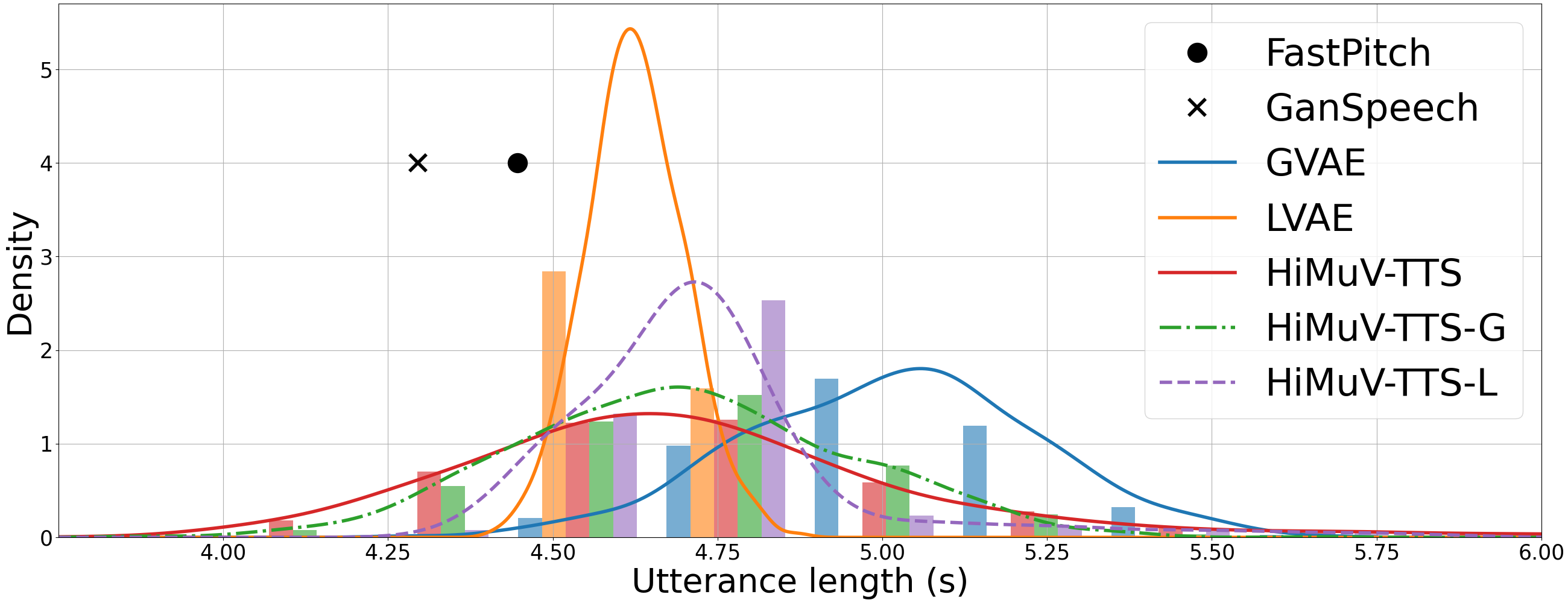}}
    \centerline{\includegraphics[width=0.8\linewidth]{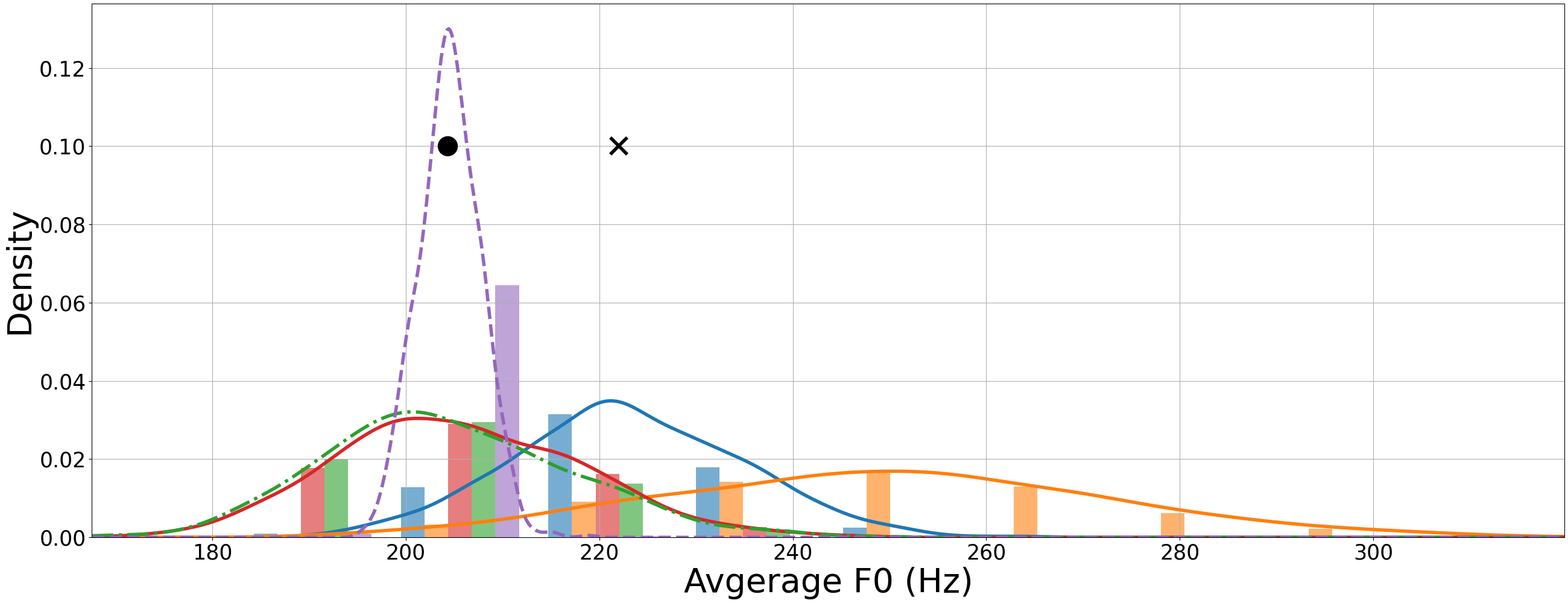}}
    \centerline{\includegraphics[width=0.8\linewidth]{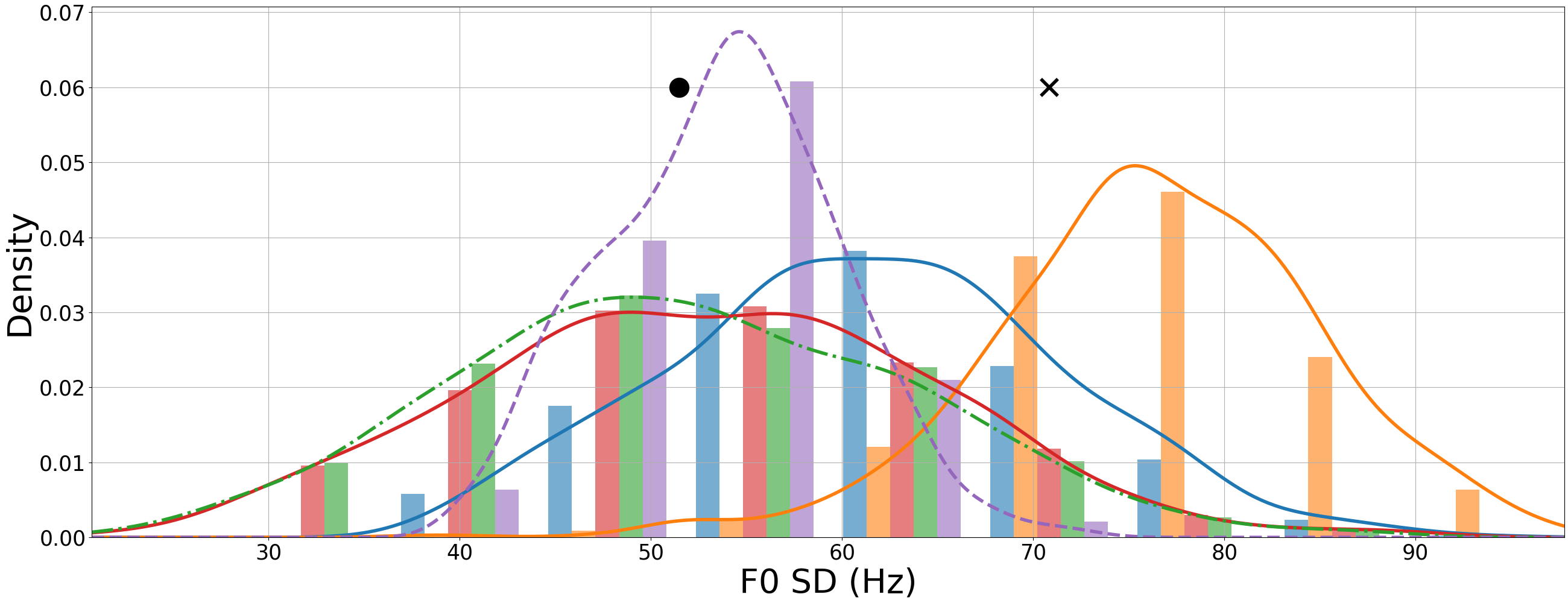}}
    \caption{Example histograms of utterance length (\textit{top}), average pitch (\textit{middle}), and pitch SD within an utterance (\textit{bottom}). 100 samples were generated from each model except the FastPich and GANSpeech with the same text and with different VAE sampling.
    Note that the FastPitch and GANSpeech models only provide a single value.}
    \vspace{-0.5cm}
\label{fig:hist}
\end{figure}

\subsection{Sample Diversity Improvement}
To verify the sample diversity improvement of the HiMuV-TTS model, we synthesized 100 different speech samples\footnote{Audio samples can be found online: \url{https://nc-ai.github.io/speech/publications/himuv-tts}} from the GVAE, LVAE, and HiMuV-TTS models for each sentence in the test dataset. The SD of the utterance length ($\sigma_{l}$), average energy of utterance ($\sigma_e$), average pitch of utterance ($\sigma_p$), and pitch SD within utterance ($\sigma_{\sigma_p}$) were computed for each sentence and averaged through the test dataset. An utterance with a low pitch SD value indicates that it is spoken in a monotonous tone. The results are summarized in Table \ref{table:result1} and the example histograms of each prosody feature for the generated speech samples from a randomly selected text sentence\footnote{The text was "They entered a stone cold room and were presently joined by the prisoner."} can be found in Figure \ref{fig:hist}. The GVAE model achieved low SD values for all prosody features, which indicates that the speech samples were generated with less diversity. The LVAE model had the highest SD value in $\sigma_p$. This was because the LVAE model focused on the local-scale prosody features, and pitch is the most locally varying prosody feature compared to others. However, this high $\sigma_p$ value of the LVAE model resulted in extremely unnatural speech (Table \ref{table:mos}), because it could not generate speech with temporally consistent pitch contour. Meanwhile, the HiMuV-TTS model could achieve high SD values generally for overall prosody features because it could well represent the various prosody attributes in a hierarchical and multi-scale way.

\subsection{Latent Representations of Each Scale}
\label{sec:exp-representation}
\begin{figure}[t]
{
    \begin{minipage}[b]{0.49\linewidth}
      \centering
        \centerline{\includegraphics[width=0.94\linewidth]{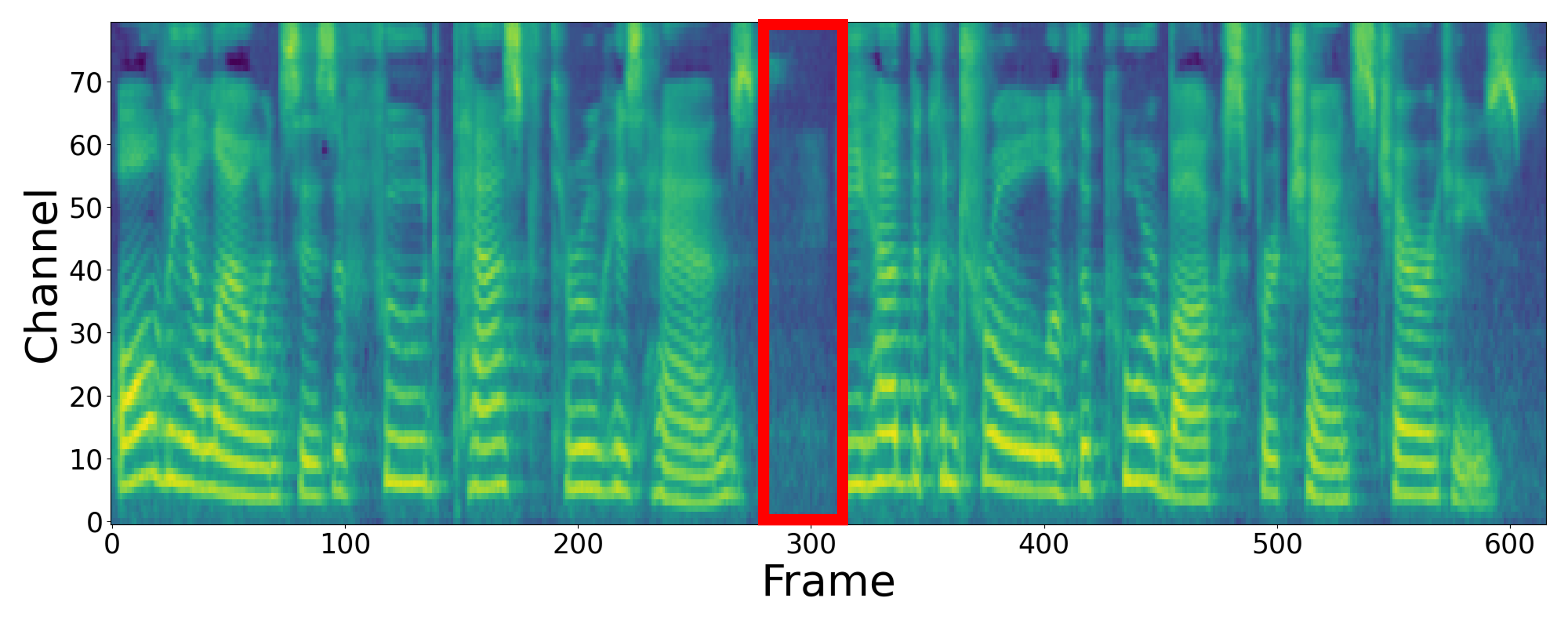}}
    \end{minipage}
    \label{fig:mel-1}
    \begin{minipage}[b]{0.49\linewidth}
      \centering
      \centerline{\includegraphics[width=0.94\linewidth]{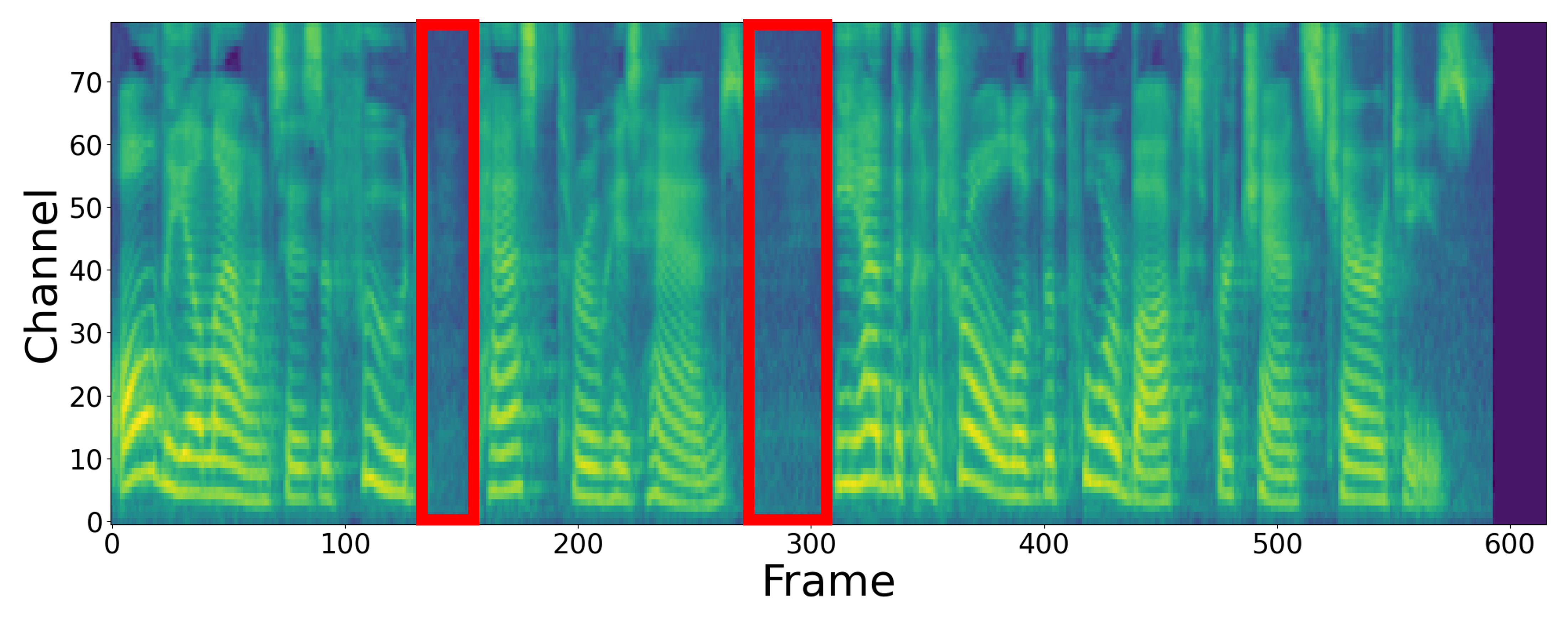}}
    \end{minipage}
    \label{fig:mel-2}
    \\
    \begin{minipage}[b]{0.49\linewidth}
      \centering
      \centerline{\includegraphics[width=0.94\linewidth]{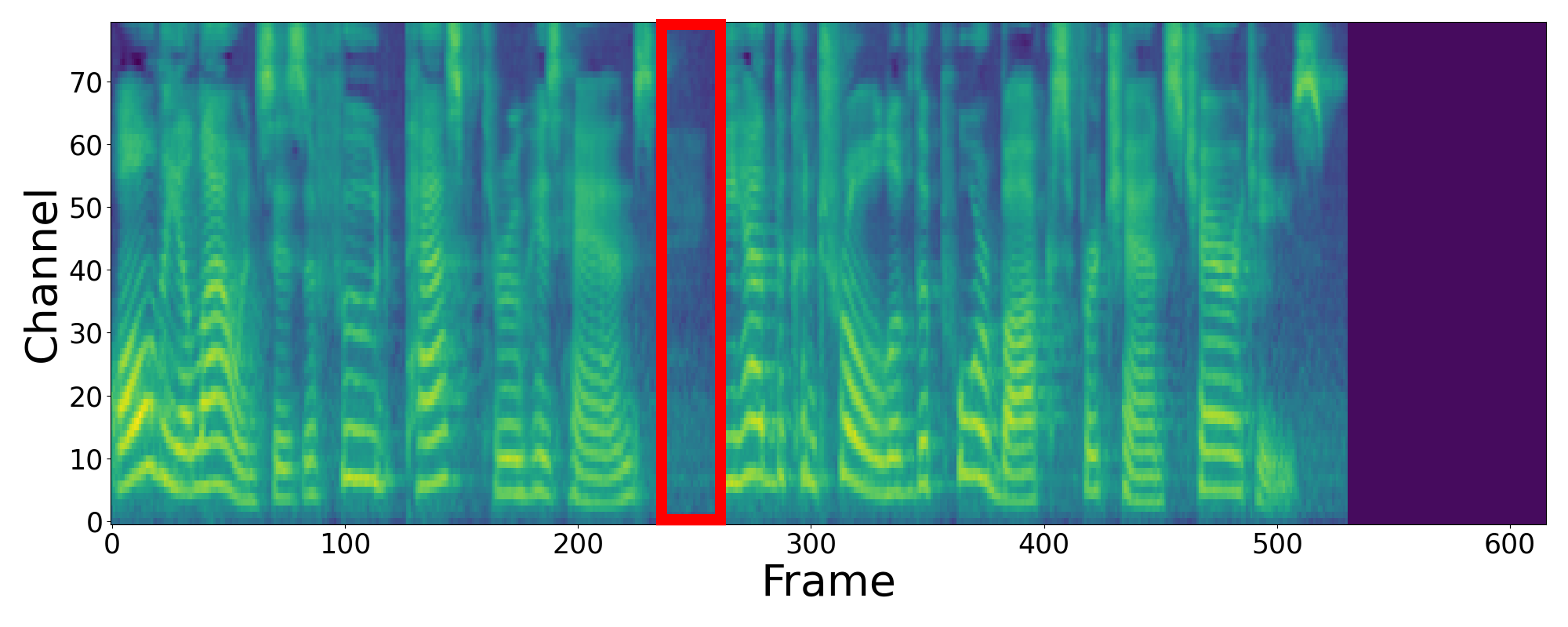}}
    \end{minipage}
    \label{fig:mel-3}
    \begin{minipage}[b]{0.49\linewidth}
      \centering
      \centerline{\includegraphics[width=0.94\linewidth]{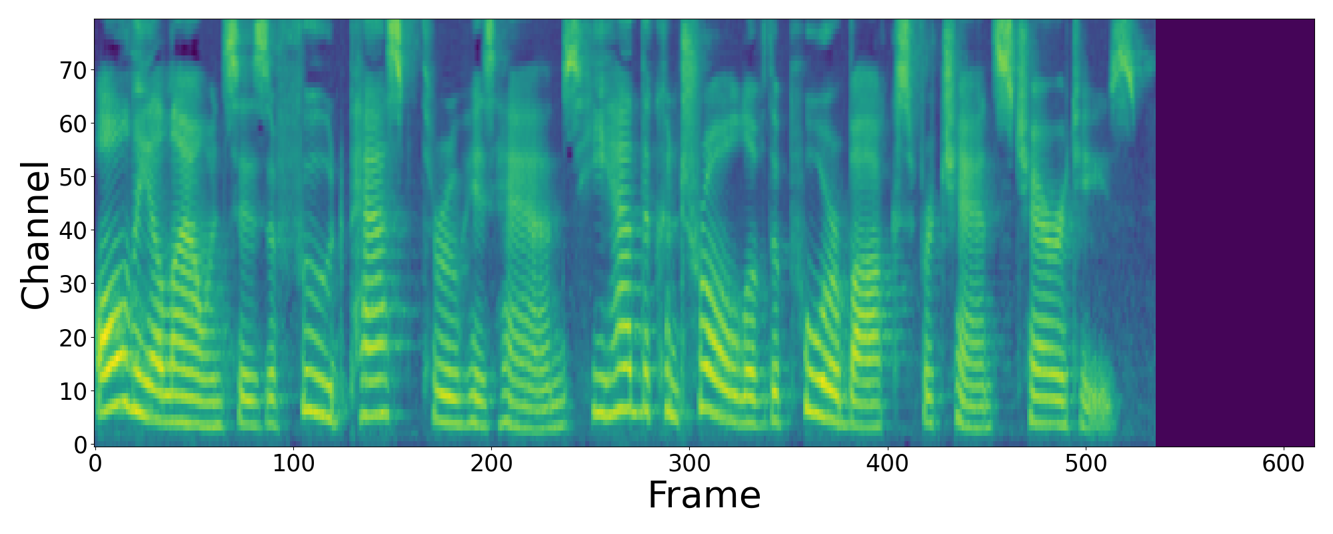}}
    \end{minipage}
    \label{fig:mel-4}
    \vspace{-0.1cm}
    \caption{Example mel-spectrograms of HiMuV-TTS-G (\textit{left column}) and HiMuV-TTS-L (\textit{right column}) with different sampling. The red boxes indicate pauses. With the HiMuV-TTS-G model, the overall speaking speed differed by sampling, while the location of pauses was preserved. Meanwhile, the HiMuV-TTS-L model varied the position and existence of pauses while maintaining the overall speaking speed.
    \label{fig:mel}
    }
}
\end{figure}

Next, we compared the HiMuV-TTS models sampled in $\bm{z}^g$ but with fixed $Z^l$ (\textbf{HiMuV-TTS-G}) and sampled in $Z^l$ but with fixed $\bm{z}^g$ (\textbf{HiMuV-TTS-L}). The results are displayed in Table \ref{table:result1} and Figure \ref{fig:hist}. Unlike the GVAE and the LVAE model, the HiMuV-TTS model modeled the utterance-level prosody attributes such as overall tone or speaking style into $\bm{z}^g$, and the remaining locally varying prosody was modeled in $Z^l$. Therefore, the HiMuV-TTS-G model had high $\sigma_p$ and $\sigma_{\sigma_p}$ values, while those of the HiMuV-TTS-L model were low. The utterance length was varied by both scales of latent variables, but was changed for different reasons. As illustrated in Figure \ref{fig:mel}, the overall speaking speed of generated speech differed by sampling in the global-scale. Meanwhile, sampling in the local-scale varied the short pauses while the overall speaking speed was preserved.

\subsection{Intelligibility}
To evaluate the intelligibility of the synthesized speech, we computed the word error rate (WER) for each model. The conformer-based speech recognition model \cite{conformer} trained with the LibriSpeech dataset \cite{librispeech} was used. The results are shown in Table \ref{table:mos}. The GANSpeech model had a significantly lower WER value compared to the FastPitch model. This was because the adversarial training technique was adopted in the GANSpeech model. The WER results of the HiMuV-TTS model were slightly higher than the GANSpeech model, because of the diverse speaking style generated by the HiMuV modules. Meanwhile, the HiMuV-TTS model with the temperature value $\tau=0$, which was multiplied to the prior SD, achieved the lowest WER results. When we analyzed the generated speech sample of the GVAE model, we observed that the speaking speed was commonly lower than the other models as shown in Figure \ref{fig:hist}. Therefore, the GVAE model had the second-lowest WER results.

\begin{table}[t]
{
    \caption{MOS test results with 95\% confidence intervals for naturalness and WER results for intelligibility evaluation.
    $\tau$ denotes the temperature value multiplied to the prior SD.}
    \vspace{-0.6cm}
    \label{table:mos}
    \begin{center}      
    \begin{tabular}{lrr}
        \toprule
        \multicolumn{1}{l}{\textbf{Model}} &
        \multicolumn{1}{c}{\textbf{MOS}} & 
        \multicolumn{1}{l}{\textbf{WER (\%)}}\\
        \midrule
        GT & $4.40 \pm 0.08$  & $8.65$ \\
        GT-Mel & \multicolumn{1}{c}{-} & $9.14$ \\
        \midrule
        FastPitch & $3.97 \pm 0.09$ & $9.77$ \\
        GANSpeech & $3.98 \pm 0.09$ & $9.38$ \\
        GVAE & $3.71 \pm 0.10$ & $9.22$ \\
        LVAE& $2.83 \pm 0.11$ & $10.34$ \\
        \midrule
        HiMuV-TTS & $3.97 \pm 0.09$ & $9.46
        $ \\
        HiMuV-TTS ($\tau$=0.0) & $\mathbf{4.04 \pm 0.09}$ & $\mathbf{9.14}$ \\
        \bottomrule
    \end{tabular}
    \end{center}
    \vspace{-0.6cm}
}
\end{table}

\subsection{Subjective Evaluation}
To evaluate the naturalness of the synthesized speech, we conducted the mean opinion score (MOS) test. Seventeen samples were generated from each model. A total of 17 native English speakers participated and scored the naturalness of the generated speech samples from 1 to 5. The results are summarized in Table \ref{table:mos}. The HiMuV-TTS model achieved similar MOS results as the FastPitch and GANSpeech models while generating speech with diver speaking styles. Furthermore, it achieved the best scores with $\tau=0$. The LVAE model had significantly low MOS results because of the inconsistent prosody of generated speech among the text sequences.

\section{Conclusion}
We propose a hierarchical and multi-scale VAE architecture to generate diverse and natural speech for the NAR-TTS model. The HiMuV-TTS model first determines the global-scale prosody. The local-scale prosody of speech is then determined via conditioning on the global-scale prosody and a learned text representation. Additionally, we adopt the adversarial training to improve speech quality. Through experiments, we showed that the proposed HiMuV-TTS model can generate speech with more naturalness and diversity as compared with the TTS models with single-scale VAEs.

\bibliographystyle{IEEEtran}

\bibliography{mybib}

\end{document}